\documentstyle[prl,aps,twocolumn,epsf,amssymb]{revtex}
\def\ave#1{\langle #1\rangle_\beta}

\def\R{{\Bbb{R}}}

\newcommand{\half}{\textstyle{\frac{1}{2}}}
\newcommand{\third}{\textstyle{\frac{1}{3}}}
\newcommand{\quar}{\textstyle{\frac{1}{4}}}

\begin{document}
\title{Momentum conservation implies anomalous 
energy transport in 1d classical lattices}
\author{Toma\v z Prosen\cite{Tomaz} and David K. Campbell\cite{David}}
\address{CNLS, MS B258, Los Alamos National Laboratory, Los Alamos, 
NM 87545}
\date{\today}
\draft
\maketitle
\begin{abstract}
Under quite general conditions, we prove that for classical many-body
lattice Hamiltonians in one dimension (1D) total momentum conservation 
implies anomalous conductivity in the sense of
the divergence of the Kubo expression for the coefficient of 
thermal conductivity, $\kappa$. Our results provide rigorous
confirmation and explanation of many of the existing
``surprising'' numerical studies of anomalous conductivity
in 1D classical lattices, including the celebrated
Fermi-Pasta-Ulam problem.

\end{abstract}

Since the pioneering work of Fermi, Pasta, and Ulam (FPU)
revealed the ``remarkable little discovery'' \cite{fpu55} that
even in strongly nonlinear one-dimensional (1D)
classical lattices recurrences of the initial
state prevented the equipartition of energy and consequent
thermalization, the related issues of thermalization, transport,
and heat conduction in 1D lattices
have been sources of continuing interest (and frustration!)
for several generations of physicists.
The complex of questions following from the FPU study involves
the interrelations among equipartition of energy (is there
equipartition? in which modes?), local thermal equilibrium
(does the system reach a well-defined temperature locally?
if so, what is it?), and transport of energy/heat
(does the system obey Fourier's heat law? If not, what
is the nature of the abnormal transport?) In sorting
through these questions, it is important to recall that
the study of heat conduction (Fourier's
heat law) is the search for a non-equilibrium steady
state in which heat flows across the system, but the
situation is usually analyzed, using the Green-Kubo formalism of linear
response \cite{kubo59}, in terms of the correlation functions in the
thermal equilibrium (grand canonical) state.
A series of reviews spread over nearly
two decades has provided snapshots of the understanding
(and confusion) at different stages of this odyssey
\cite{toda75,visscher76,macdonald78,jackson78,toda79,ford92}.

Much of the past effort has been devoted to
attempts to verify Fourier's law of heat conduction
\begin{equation}
\langle \vec{J}\rangle = \kappa \nabla T,
\label{fourier;eq}
\end{equation}
where in 1D the gradient is replaced by the derivative with
respect to $x$.
Here, $\kappa$ is the transport coefficient of thermal conductivity.
Strictly speaking, $\kappa$ is well defined only for a system that
obeys Fourier's law and where a {\em linear} temperature gradient is 
established (for small energy gradients such that relative temperature
variation across the chain is small; in general $\kappa$ is a function
of temperature, of course). In the literature the dependence
of $\kappa(L)$ on the size $L$ of the system/chain has also been used 
to characterize the (degree of) anomalous transport.
However, the definition of $\kappa$ for an anomalous conductor, where no
internal temperature gradient may be established, is
ambiguous. Typically, one defines it in the ``global''
sense, as  $\kappa(L) \equiv \kappa_{G} \equiv J L/\Delta T$,
where $\Delta T$ is the total temperature difference
between the two thermal
baths. 
However, if the temperature gradient is {\it not} constant across the system, 
and/or if there are finite temperature gaps between the thermal baths
and the edges of the system due to system-bath contact,
one should define and study a local $\kappa$,
$\kappa \equiv \kappa_{L} \equiv {J \over \nabla T}$, where
$ \nabla T$ is the {\it local} thermal gradient. 
A very wide range of results have been produced
by previous studies of different systems:
\begin{itemize}
\item in acoustic harmonic chains, rigorous results
  \cite{rieder67}, establish that no thermal gradient
can be formed in the system, with the result that
formally  $\kappa_{G} \sim L^1$,
which can be understood
heuristically by the stability of the linear Fourier modes
and the absence of mode-mode coupling; 
\item in the ``Toda lattice,'' an integrable lattice model
  \cite{toda75,toda67}, in which the result
  $\kappa_{G} \sim L^1$ \cite{expL} can be understood in terms
  of stable, uncoupled {\it nonlinear} modes, the solitons,
  which are a consequence of the system's
complete integrability \cite{toda79};
\item in non-integrable models with smooth potentials,
  including (i) the FPU system, leading eventually to claim that
  chaos was necessary and sufficient for normal
  conductivity ($\kappa_{G}= \kappa_{L} \sim L^0$) \cite{ford92},
  a claim that has been countered by convincing
  numerical evidence for anomalous conductivity
  in FPU chains ($\kappa_{L} \sim L^{0.4}$)
  \cite{kaburaki93,lepri97}; (ii) the diatomic (and
   hence non-integrable) Toda lattice, where
   initial numerical results claiming
   $\kappa_{L} \sim L^0$ \cite{jackson89} have recently
   been refuted by a more systematic study showing
   $\kappa_{L} \sim L^{0.4}$ \cite{hatano99}; and (iii)
   the ``Frenkel-Kontorova model,'' where recent studies
   have shown that (at least for low temperatures)
   $\kappa_{L} \sim L^0$ \cite{hu98}; 
\item in non-integrable models with hard-core potentials,
  including (i) the ``ding-a-ling'' model \cite{casati84}; 
  (ii) the ``ding-dong'' model \cite{prosen92}, 
and (iii) even simpler single particle chaotic billiard 
model \cite{alonso}, where numerics
  show convincingly that $\kappa_{L} = \kappa_{G} \sim L^0$.
\end{itemize}

This bewildering array of results has 
recently been partially clarified in a series
of independent but overlapping studies.
The numerical studies of Hu {\it et al.} \cite{hu98} and of
Hatano \cite{hatano99} show that {\it overall momentum conservation}
appears to a key factor in anomalous transport in 1D lattices.
Lepri {\it et al.} \cite{lepri98a,lepri98b} and
Hatano\cite{hatano99} have argued
that the anomalous transport in momentum conserving
systems can be understood in terms of low frequency,
long-wavelength ``hydrodynamic modes'' that
exist in typical momentum conserving systems and 
that hydrodynamic arguments may explain
the exponents observed in FPU \cite{lepri98a,lepri98b} and
diatomic Toda lattice \cite{hatano99}.

In the present Letter, we extend and formalize these recent results
and resolve finally at least
one important aspect of conductivity in 1D lattices: namely,
we present a rigorous proof that in 1D {\it conservation
of total momentum implies anomalous conductivity} 
provided only that the average pressure is non-vanishing in
thermodynamic limit.

We consider the general class of classical 1D many-body
Hamiltonians
\begin{equation}
H=\sum_{n=0}^{N-1}\left(\frac{1}{2 m_n} p_n^2 + V_{n+1/2}(q_{n+1} -
  q_n)\right) ,
\label{eq:ham}
\end{equation}
where $V_{n+1/2}(q)$ is an arbitrary (generally non-linear) interparticle
interaction. Note that the potential, $V_{n+1/2}$,
depends only on the {\it differences} between two adjacent sites; in
particular, there
is {\it no} ``on-site'' potential, $U_{\rm OS}(q_n)$, depending
on the individual coordinates.
The (finite) system is considered to be defined on a
system
of length $L =Na$ with periodic boundary conditions 
$(q_N,p_N) \equiv (q_0,p_0)$, where the actual particle positions
are $x_n = n a + q_n$.
In our analysis the masses $m_n$, 
as well as interparticle potentials $V_{n+1/2}(q)$, can have 
{\em arbitrary dependence on the sites} $n$,
though the examples studied in literature to date have mostly had
uniform potentials $V_{n+1/2}(q)=V(q)$ and uniform, $m_n=m$, or dimerized
$m_{2n}=m_1, m_{2n+1}=m_2$, masses.
We require only that the 
Hamiltonian (\ref{eq:ham}) be invariant under translations
$q_n \rightarrow q_n + b$ for {\it arbitrary} $b$. This requires
$U_{\rm OS}(q_n) = 0$ \cite{hu98}. Note that we may write
the Hamiltonian in Eq.(\ref{eq:ham}) as
$H=\sum_{n=0}^{N-1} h_{n+1/2}$, where $h_{n+1/2}$ is
the Hamiltonian density
\begin{equation}
h_{n+1/2} = \frac{p^2_{n+1}}{4 m_{n+1}} + \frac{p^2_n}{4 m_n}
+ V_{n+1/2}(q_{n+1}-q_n),
\label{eq:hamdens}
\end{equation}

Our aim is to estimate $\kappa$, the coefficient
of thermal conductivity, which is given by the Kubo formula
\cite{linearresp}
\begin{equation}
\kappa = \lim_{T\rightarrow\infty}\lim_{L\rightarrow\infty} 
\frac{\beta}{L}
\int_{-T}^T dt \ave{J(t)J}.
\label{eq:Kubo}
\end{equation}
Here we have written the canonical average of an observable $A$ at inverse
temperature $\beta$ as
$\ave{A} = \int \Pi_n dp_n dq_n A \exp(-\beta H)
/\int \Pi_n dp_n dq_n \exp(-\beta H)$. The order
of limits in Eq. (\ref{eq:Kubo}) is crucial to the precise
definition of $\kappa$ \cite{kubo59}. In Eq.(\ref{eq:Kubo}),
$J=\sum_{n=0}^{N-1} j_n$ is the total heat current, and
$j_n$ is the heat current density \cite{hu98}, given by
\begin{eqnarray}
j_n &=& \{ h_{n+1/2},h_{n-1/2} \} = \label{eq:currdens} \\
&=& \frac{p_n}{2 m_n}\left(V_{n+1/2}'(q_{n+1}-q_n) + 
V_{n-1/2}'(q_n-q_{n-1})\right),
\nonumber
\end{eqnarray}
where $\{,\}$ is the usual canonical Poisson bracket.

Using Eq.s (\ref{eq:hamdens},\ref{eq:currdens}),
we find that current density given by (\ref{eq:currdens}) 
satisfies the continuity equation
\begin{equation}
\dot{h}_{n+1/2} = \{ H,h_{n+1/2} \} = j_{n+1} - j_n.
\end{equation}

Our ensuing analysis is similar to that used by Mazur 
\cite{mazur69}, with a crucial difference: we will average
correlation functions over a {\em finite} rather than {\em infinite} time 
domain, $T$. We start with an elementary inequality.
For an arbitrary observable $X(t) = X(\{q_n(t),p_n(t)\})$, we have
\begin{equation}
\int_{-\infty}^\infty dt g_T(t) \ave{X(t)X} \ge 0 
\label{eq:basineq}
\end{equation}
where $g_T(t)$ is a suitable $L^2(\R)$ {\em window} function of
effective width $T$,
which has the following properties:\\\\
(i) $\int_{-\infty}^\infty dt g_T(t) = T$.\\
(ii) $\int_{-\infty}^\infty dt g^2_T(t) = T$.\\
(iii) $\tilde{g}(\omega) :=
\int_{-\infty}^\infty dt g_T(t) e^{i\omega t} > 0$ for all $\omega$.\\\\
The natural choice satisfying these conditions is
a Gaussian, $g_T(t) = \sqrt{2}\exp(-2\pi (t/T)^2)$.
Using elementary Fourier analysis, the above inequality (\ref{eq:basineq})
is easily proved by rewriting it as
\begin{equation}
\int d\omega \tilde{g}_T(\omega) \ave{S_X(\omega)} \ge 0
\label{eq:basineq2}
\end{equation}
where $S_X(\omega) = 
\lim_{T\rightarrow\infty}\frac{1}{T}
\left|\int_0^T dt e^{i\omega t} X(t)\right|^2$ 
is the power spectrum of the signal $X(t)$. Obviously, $S_X(\omega)>0$, and
given (iii), the inequality (\ref{eq:basineq},\ref{eq:basineq2}) is
clearly fulfilled. 
We now write the observable $X$ as $X=A + \alpha B$, $\alpha\in\R$.
Optimizing with respect to the
parameter $\alpha$, we arrive at the Schwartz-like inequality
\begin{eqnarray}
&&\left(\int dt g_T(t)\ave{A(t)A}\right)
\left(\int dt g_T(t)\ave{B(t)B}\right) \ge \nonumber \\
&& \ge \left(\int dt g_T(t)\ave{B(t)A} \right)^2. \label{eq:schwartz}
\end{eqnarray}
The above inequality is of quite general use. We implement
it by taking $A \equiv J$ and $B \equiv P$,
where $P=\sum_{n=0}^{N-1} p_n$ is the
total momentum. For Hamiltonians of the form (\ref{eq:ham}),
$P$ is an integral of motion $\dot{P}=\{H,P\}\equiv
0$ due 
to translational symmetry.
Since $P(t)=P$, the inequality (\ref{eq:schwartz}) reads
\begin{eqnarray}
\int dt g_T(t) \ave{J(t) J} \ge T \frac{\ave{J P}^2}{\ave{P^2}}
\label{eq:JJineq}.
\end{eqnarray}

The RHS of Eq.(\ref{eq:JJineq}) can be easily evaluated:
$\ave{P^2} = N/\beta$,
and $\ave{J P} = \beta^{-1}\sum_{n=0}^{N-1} \ave{V'(q_{n+1}-q_n)}$
since we have in general that 
$\ave{A(\{q_n\})B(\{p_n\})}=\ave{A(\{q_n\})}\ave{B(\{q_n\})}$.
$\ave{V'_{n+1/2}(q_{n+1}-q_n)}$ is an average {\em force} between 
particles $n$ and $n+1$, i.e. {\em the thermodynamic pressure}
and does not depend on $n$. (In thermal equilibrium the average net
force on particle $n$ vanishes and hence
$\ave{V'_{n-1/2}(q_{n}-q_{n-1})} = \ave{V'_{n+1/2}(q_{n+1}-q_{n})}$.)
The pressure can be rewritten 
through the usual thermodynamic definition 
$$\phi \equiv \frac{\partial F}{\partial L}
=\frac{1}{N}\frac{\partial F}{\partial a}=\frac{1}{L}\sum\limits_{n=0}^{N-1} 
\ave{V'_{n+1/2}(x_{n+1}-x_n+a)},$$ 
where $\exp(-\beta F)=\int\Pi_n dp_n dq_n \exp(-\beta H)$.
Inserting the above and multiplying with $\beta/L$, we find that
inequality (\ref{eq:schwartz}) reads
\begin{equation}
\frac{\beta}{L}\int dt g_T(t)\ave{J(t)J} \ge T \phi^2.
\label{eq:result}
\end{equation}
Since the Kubo formula can be equivalently written in terms of the 
window function as
$\kappa = 2^{-1/2} \lim_{T\rightarrow\infty}\int dt g_T(t) C(t)$,
where $C(t)=\lim_{L\rightarrow\infty}\ave{J(t)J}/L$, since
$\lim_{T\rightarrow\infty}g_T(t) = \sqrt{2}$, and implementing
in the above result
(\ref{eq:result}) the two limits as indicated in (\ref{eq:Kubo}),
we have proved 
our main result:\\\\
{\bf Theorem:} In momentum conserving systems of type (\ref{eq:ham}),
if the pressure is non-vanishing in the thermodynamic limit, 
$\lim_{L\rightarrow\infty}\phi > 0$, then
the thermal conductivity diverges and $\kappa \rightarrow \infty$.
\\\\
Therefore, we find anomalous energy transport as a simple consequence
of the total momentum conservation.
The only case in which the pressure is expected to vanish at {\em any
temperature} is when the forces between particles at zero
temperature equilibrium are zero ($V'_{n+1/2}(0)=0$) {\it and}
the interparticle potentials are all even functions 
($V_{n+1/2}(q)=V_{n+1/2}(-q)$) so that the forces are also expected (and
found numerically for $\beta$ FPU problem) to average
to zero for arbitrary canonical thermal fluctuations. 
This is indeed the case for the $\beta$ FPU 
problem, where $V_{n+1/2}(q) = \half q^2 + \quar \beta q^4$
\cite{lepri97,lepri98b}, and 
there the integrated correlation function diverges for more subtle
(dynamical) reasons (the slow asymptotic power-law decay of current-current 
correlation function $\sim t^{-0.6}$).

Even if the zero temperature equilibrium forces vanish
$V'_{n+1/2}(0)=0$, we still have non-vanishing finite temperature
pressure (due to `thermal expansion' of a system confined to a 
fixed volume $L=a N$) whenever interparticle potentials are not even.
This is the case for the $\alpha$ FPU model, $V_{n+1/2}(q)=\half q^2 +
\third \alpha q^3$, for the modified diatomic Toda lattice \cite{hatano99}, 
$V_{n+1/2}(q) = \exp(-q) + q$, and for the diatomic hard-point
1D gas \cite{casati85,hatano99} $V_{n+1/2}(q) = 
\{0 \;\;{\rm if}\;  q > -a; \infty \;\;{\rm if}\; q \le -a \}.$
For the usual diatomic Toda lattice \cite{hatano99},
$V_{n+1/2}=\exp(-q)$ the pressure is non-vanishing even at zero
temperature, since $V'_{n+1/2}(0)\neq 0$.

To augment and illustrate our analytic discussion, we have simulated
numerically the current-current
autocorrelation function $\ave{J(t)J}/L$ in a generic anharmonic
chain, namely in the ``$\alpha\beta$'' FPU model
with $V_{n+1/2}(q) = \half q^2 + \third \alpha q^3
+ \quar \beta q^4$ where we take $\alpha=2,\beta=4$ and energy per
particle $E/N=1$.
In Fig. 1 we compare the results for $N=16,32,64$ with the
equillibrium value of
the squared pressure $\phi^2 = 0.964\ldots$.
We have also checked numerically that for the symmetric interparticle
potential (same as above except with $\alpha=0$) the pressure
indeed vanishes and 
the current-current correlation functions decay
asymptotically as $\sim t^{-0.6}$
(in agreement with results of Refs.\cite{lepri98a,lepri98b}).
  
Given that momentum conservation implies anomalous
conductivity, it is natural to ask whether the converse is true: namely,
does anomalous conductivity imply that the model conserves momentum?
Two counterexamples show that this result is {\it not} true.
First, if one considers a {\it linear} chain of {\it optical}
phonons---so $ V_{n+1/2} \sim (q_{n+1} - q_n)^2$ and
$U_{\rm OS} \sim q_n^2$---one can show \cite{prosen99} by a
straightforward extension of the arguments of Ref.\onlinecite{rieder67}
that this momentum non-conserving model nonetheless has anomalous
transport. Similarly, there is a momentum {\it non-conserving}
integrable model due to Izergin and Korepin \cite{izergin81} that
also shows anomalous conductivity \cite{prosen99}.
Finally, let us to stress that in 1D lattices the nature of
dynamics, whether it be completely integrable, completely
chaotic, or mixed, does not affect our result: if
total momentum is conserved and the canonical average of the pressure
does not vanish, the transport is anomalous. We shall address
the central issue of the necessary and sufficient conditions for
normal transport in a forthcoming paper \cite{prosen99}.

The authors gratefully acknowledge the hospitality of Center for 
Nonlinear Studies (CNLS), Los Alamos National Laboratory,
where the work was performed. TP acknowledges financial support by the
Ministry of Science and Technology of the Republic
of Slovenia, and DKC thanks the CNLS supporting him as
the 1998-99 Ulam Scholar. This research was supported
in part by the Department of Energy under contract W-7405-ENG-36 and
by the National Science Foundation under grant DMR-97-12765.

\begin{figure}
\begin{center}
\leavevmode
\epsfxsize=3.5in
\epsfbox{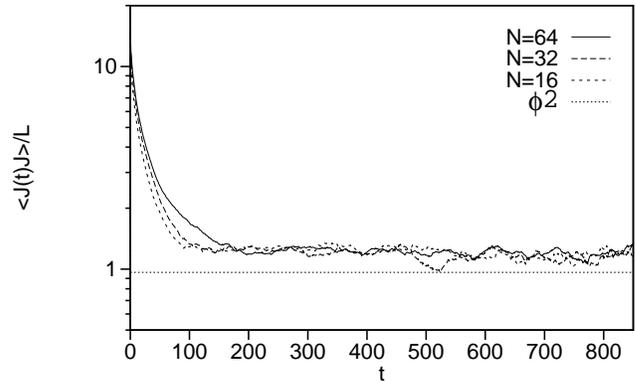}
\end{center}
\protect\caption{Current-current autocorrelation function in the
  ``$\alpha\beta$'' FPU model with $\alpha=2,\beta=4$, and $E/N=1$. We show
numerical data canonically averaged over 500 pseudo-random initial conditions 
for three different sizes $N=16,32,64$ and compare it to the 
squared pressure $\phi^2$ (dotted line).}
\label{fig}
\end{figure}


\begin{references}
\bibitem{expL} Note that in linear/integrable systems the 
temperature profile inside the system is nearly flat and almost
the entire drop in temperature occurs near the edges. In
such cases, one would find for the local thermal conductivity
$\kappa_L \sim \exp({\rm const} \times L)$.
\bibitem{linearresp} The validity of the linear response theory that
  underlies the Kubo formula is not an issue here, since
  we will be making analytic estimates that are valid for arbitrarily
  small temperature differences and since we take care to isolate
  the pathological cases of zero internal thermal gradients.
\end{references}
\end{document}